# The contribution of microbunching instability to solar flare emission in the GHz to THz range of frequencies

J. Michael Klopf[1], Pierre Kaufmann[2,3], Jean-Pierre Raulin[2] and Sérgio Szpigel[2]

[1]College of William & Mary, Dept. of Applied Science, McGlothlin-Street Hall, Williamsburg, VA 23187 USA.

[2]Centro de Rádio-Astronomia e Astrofísica Mackenzie, Escola de Engenharia, Universidade Presbiteriana Mackenzie, Rua Consolação 896, 01302-907 São Paulo, SP, Brazil.

[3]Centro de Componentes Semicondutores, Universidade Estadual de Campinas, Cidade Universitária "Zeferino Vaz", 13981-970 Campinas, SP, Brazil.

**Abstract**

Recent solar flare observations in the sub-THz range have provided evidence of a new spectral component with fluxes increasing for larger frequencies, separated from the well-known microwave emission that maximizes in the GHz range. Suggested interpretations explain the THz spectral component, but do not account for the simultaneous microwave component. We present a mechanism for producing the observed "double-spectra". Based on coherent enhancement of synchrotron emission at long wavelengths in laboratory accelerators, we consider how similar processes may occur within a solar flare. The instability known as microbunching arises from perturbations that produce electron beam density modulations, giving rise to broadband coherent synchrotron emission at wavelengths comparable to the characteristic size of the microbunch structure. The spectral intensity of this coherent synchrotron radiation (CSR) can far exceed that of the incoherent synchrotron radiation (ISR), which peaks at higher frequency, thus producing a double-peaked spectrum. Successful CSR simulations are shown to fit actual burst spectral observations, using typical flaring physical parameters and power-law energy distributions for the accelerated electrons. The simulations consider an energy threshold below which microbunching is not possible because of Coulomb repulsion. Only a small fraction of the radiating charges accelerated to energies above the threshold is required to produce the microwave component observed for several events. The ISR/CSR mechanism can occur together with other emission processes producing the microwave component. It may bring an important contribution at microwaves at least for certain events where physical conditions for the occurrence of the ISR/CSR microbunching mechanism are possible.

**Key words:** Sun: flares, coherent synchrotron radiation, microbunching instability, THZ flare emission

## I. Introduction

A number of solar bursts have been observed to have unexpected distinct spectral components in the GHz to sub-THz range: one corresponds to the well known microwave emission maximizing at few to tens of GHz, and another with fluxes increasing for larger sub-THz frequencies. Early solar burst observations made up to 0.1

THz have suggested high-frequency "double-spectral" features (Shimabukuro 1970; Croom 1971; Akabane et al. 1973; Zirin & Tanaka 1973; Roy 1979; Kaufmann et al. 1985; White et al. 1992). Observations carried out at higher frequencies (0.2 and 0.4 THz) by the Solar Submillimeter Telescope (SST), have clearly evidenced the sub-THz flux component increasing with frequency (Kaufmann et al. 2004; 2009; 2011; Silva et al. 2007). The effect has also been reported as an up-turn of the spectral trend in that range of frequencies during certain phases of other events (Raulin et al. 2004; Lüthi et al. 2004; Trottet et al. 2011). A dramatic example of a "double-spectral" structure feature was observed during an intense 30 THz impulsive burst with flux several times larger than the microwave component (Kaufmann et al. 2013).

These results raise serious problems to explain the simultaneous presence of the sub-THz and the concurrent microwave component. A number of emission processes invoked to explain the sub-THz spectral component include emission by free-free collisions of thermal electrons, synchrotron produced by high-energy electrons (Kaufmann et al. 2004; Silva el al. 2007) and by relativistic positrons (Silva et al. 2007; Trottet et al. 2008), emission by Langmuir waves excited by beams of electrons and protons at denser regions of the solar active centers (Sakai et al. 2006; Sakai & Nagasuchi 2007) and inverse-Compton effect on the field of synchrotron electrons (Kaufmann et al. 1986). Several possible mechanisms were recently reviewed by Fleishman and Kontar (2010) who have added two other possibilities: the inverse-Compton effect on field of photons produced by Langmuir waves and the Vavilov-Cherenkov emission by high-energy electrons on an assumed partially ionized chromospheric gas. The authors concluded that more than one mechanism is likely to be acting at the same time and that a free-free contribution might always be present to a certain level.

While these explanations are used to explain the sub-THz component, they however do not account for the concurrent microwaves spectral component that is also observed. One explanation might assume arbitrarily that distinct populations of electrons are accelerated at about the same time, at different energies. This assumption is often adopted to explain complex structures in radio spectra of quasars (Kellermann & Pauliny-Toth 1969). Possible supporting observational evidences that this could happen in a solar flare accelerator are the bursts with double power-law photon spectral indices of X- and gamma rays emitted by bremsstrahlung (Kurt et al. 2010) reminding that the photon spectral energy indices are directly related to the injected electrons energy spectra (Tandberg-Hanssen & Emslie 1988). Another possibility assumes a peculiar acceleration site scenario where a single beam of electrons is injected from a low altitude into two different magnetic loops, emitting microwave synchrotron radiation in the higher arch with weaker magnetic field, and sub-THz synchrotron in the lower and stronger magnetic field arch (Silva et al. 2007). This magnetic topology demands the assumption of critical selection of parameters close to limiting physical conditions at the flaring site, marginally suggested by certain observations.

In this study we examine in detail how both spectral components can be produced by a single beam of high-energy electrons undergoing physical processes similar to those occurring in laboratory accelerators as it has been recently suggested (Kaufmann & Raulin 2006; Klopf, Kaufmann & Raulin 2010). The THz spectral component may be produced by one of the above suggested mechanisms. All of these mechanisms depend on the acceleration of high-energy electrons at the early phase of the process. The simpler assumption is that incoherent synchrotron radiation (ISR) is produced by a beam of very high-energy electrons (> MeVs) with flux maximizing somewhere in the THz range of frequencies. Under the proper conditions, physical perturbations such as magnetic field small scale spatial structures or wave-particle interactions may produce modulations of the accelerated electron beams in the form of microbunching (Byrd et al. 2002; Carr et al. 2002; Nodvick & Saxon 1954; Stupakov & Heifets 2002; Venturini & Warnock 2002). As described in this paper, the emission from these modulations could produce extremely bright broadband coherent synchrotron radiation (CSR) in the GHz frequency range and thus bring a simultaneous contribution to the low-frequency spectral component, allowing explanation of the observed "double-peaked" flare emissions.

## 2. Double Spectrum Flare Observation

The first example of a two spectral component solar burst in the GHz to sub-THz range of frequencies has been obtained for the large solar flare of November 4, 2003 by the SST (0.2 and 0.4 THz) and by the Owens Valley Solar Array (0.5 - 18 GHz) and Itapetinga 13.7-m radio telescope (44 GHz only for time structure P4) (Kaufmann et al. 2004). Figure 1 shows the time profiles (a) and spectra (b) for mean fluxes of the time structures $P_1$ and $P_4$. Several events have been reported exhibiting similar spectral trends, as quoted in the previous section. The solar event of November 4, 2003 has been observed under particularly good atmosphere propagation conditions and has been selected for comparison to the laboratory accelerator ISR/CSR mechanism.

## 3. CSR in Laboratory Accelerators

Laboratory based accelerators have been developed over several decades for producing extremely bright photon beams, most commonly in the form of incoherent synchrotron radiation (ISR). Many techniques and devices have been developed to further enhance the brightness of the photon beams such as the use of periodic magnetic structures known as insertion devices (typically classified as either an undulator or wiggler) (Motz 1951; Motz & Walsh 1962; Friedman & Herndon 1973). Still further advances in accelerator technology has enabled devices known as Free Electron Lasers (FEL), which produce narrow band fully coherent photon beams with unparalleled brightness (Colson 1976). In these types of devices, feedback between the insertion device, the radiation field, and the electron beam results in a modulation of the energy and spatial distribution of the electrons, known as microbunching. In addition, to achieve maximum brightness and full longitudinal coherence in many FEL devices, the electron beam is compressed into very short bunches before passing through the insertion device. Though this modulation and bunching of the electron beam is typically carefully controlled, instability conditions have also been demonstrated that give rise to spontaneous microbunching of the electron beam (Byrd et al. 2002; Carr et al. 2002; Stupakov & Heifets 2002; Venturini & Warnock 2002).

The bunching or microbunching of the electron beam, by any of these means, results in a coherent enhancement of the synchrotron radiation at wavelengths comparable to, or longer than the characteristic length scale of the bunch structure (Nodvick & Saxon 1954). The spectral emission of the bunched beams in many of these laboratory accelerators and the dynamics of the accelerated electrons may exhibit some shared physics with the solar flare emissions described in the previous section. In particular, we examine here the process of broadband CSR, which has been shown to very efficiently produce intense radiation at wavelengths longer than the characteristic bunch/microbunch size (Williams 2002, Byrd et al. 2002, Carr et al. 2002).

To understand the CSR process, we first recall that when relativistic electrons are accelerated in a dipole magnetic field, they emit synchrotron radiation. At wavelengths short compared to the size of the electron bunch (or microbunch structure), the emission from the electrons is incoherent and the resulting radiation exhibits the well-known ISR spectrum emitted by charged particles accelerated to relativistic energies (Schwinger 1949; Ginzburg & Syrovatskii 1965). For emission at wavelengths approximately equal to or longer than the size of the bunch/microbunch, the near field of the radiation from each electron overlaps the entire bunch structure, resulting in a coherent interaction and a spectral brightness that scales as the square of the number of electrons within the bunched region (Nodvick & Saxon 1954). The coherent interaction makes the CSR emission extremely efficient and has been suggested as the highest average power terrestrial THz source (Carr et al. 2002).

The spectrum of synchrotron radiation emitted by an electron bunch is derived by generalizing the classical theory of electrodynamics for a single radiating electron (Jackson 1998) to a system with multiple electrons (Williams et al. 1989; Hirschmugl et al. 1991; Hulbert et al. 2001; Carr et al. 2002; Williams 2002; 2006). The so called Liénard-Wiechert electric field in frequency-domain for a single radiating electron (in Gaussian cgs units) is given by:

$$\vec{E}(\omega) = \frac{e}{c}\int_{-\infty}^{\infty} \frac{\hat{n}\times[(\hat{n}-\vec{\beta}_e)\times\dot{\vec{\beta}}_e] + cR^{-1}(\tau)\gamma^{-2}(\hat{n}-\vec{\beta}_e)}{(1-\hat{n}\cdot\vec{\beta}_e)^2 R(\tau)} \exp[i\omega(\tau + R(\tau)/c]\, d\tau \qquad (1)$$

where $e$ is the electron charge, $c$ is the velocity of light, $\vec{\beta}$ is the ratio of the velocity of the electron to the velocity of light, $\dot{\vec{\beta}}$ is the acceleration of the electron divided by c, $\gamma$ is the ratio of the mass of the electron to its rest mass $m_e c^2$, $R(\tau) = |\vec{R}(\tau)| = |\vec{x}-\vec{r}(\tau)|$ is the distance from the position of the radiating electron $\vec{r}(\tau)$ relative to an origin $O$ at the retarded time $\tau$ to the position of the observation point $\vec{x}$ relative to $O$ at time $t = \tau + R(\tau)/c$ and $\hat{n}$ is a unit vector in the direction of $\vec{x}-\vec{r}(\tau)$ (note that the integration is over the retarded time $\tau$).

One should note that Eq. (1) includes both the far-field term ("acceleration field"), which depends linearly on the acceleration of the electron $\dot{\vec{\beta}}$, and the near-field term ("velocity field"), which is independent of the acceleration. Normally, the near-field term is not considered in synchrotron calculations, but in the case of far infrared synchrotron radiation, particularly at THz frequencies and below, the contribution of such a term is significant and should be included (Williams 2006). Assuming that the observation point is far from the acceleration region we can apply a far-field approximation. In this case, the unit vector $\hat{n}$ can be considered as nearly constant in time and the distance $R(\tau)$ can be approximated by

$$R(\tau) \approx x - \hat{n}\cdot\vec{r}(\tau) \ . \qquad (2)$$

Then, the single-particle intensity $I_e(\omega)$ (energy radiated per unit of solid angle per unit of angular frequency interval in Gaussian cgs units) for a radiating electron is given by:

$$I_e(\omega) = \frac{d^2W_e}{d\omega d\Omega} = \frac{e^2}{4\pi^2 c} \left| \int_{-\infty}^{\infty} \frac{\hat{n} \times [(\hat{n} - \vec{\beta}_e) \times \dot{\vec{\beta}}_e]}{(1 - \hat{n} \cdot \vec{\beta}_e)^2} \exp[i\omega(\tau - \hat{n} \cdot \vec{r}(\tau)/c)] d\tau \right|^2 . \tag{3}$$

Using the relation

$$\frac{\hat{n} \times [(\hat{n} - \vec{\beta}_e) \times \dot{\vec{\beta}}_e]}{(1 - \hat{n} \cdot \vec{\beta}_e)^2} = \frac{d}{dt}\left[ \frac{\hat{n} \times (\hat{n} \times \vec{\beta}_e)}{1 - \hat{n} \cdot \vec{\beta}_e} \right] , \tag{4}$$

a simpler expression for $I_e(\omega)$ can be obtained through an integration by parts:

$$I_e(\omega) = \frac{e^2 \omega^2}{4\pi^2 c} \left| \int_{-\infty}^{\infty} \hat{n} \times (\hat{n} \times \vec{\beta}_e) \exp[i\omega(\tau - \hat{n} \cdot \vec{r}(\tau)/c)] d\tau \right|^2 . \tag{5}$$

Without loss of generality, one can choose a coordinate system in which the trajectory of the radiating electron lies in the *x-y* plane with instantaneous radius of curvature $\rho$ and the unit vector $\hat{n}$ lies in the *x-z* plane making an angle $\theta$ with the *x* axis. We consider that at time $\tau = 0$, the electron is situated at the origin $O$. For small angles $\theta$ and relatively short time intervals, the single-particle intensity $I_e(\omega)$ is then given by:

$$I_e(\omega) = \frac{e^2}{3\pi^2 c} \left( \frac{\omega \rho}{c} \right)^2 \left( \frac{1}{\gamma^2} + \theta^2 \right)^2 \left[ K_{2/3}^2(\xi) + \frac{\theta^2}{(1/\gamma^2) + \theta^2} K_{1/3}^2(\xi) \right] , \tag{6}$$

where $K_{2/3}(\xi)$ and $K_{1/3}(\xi)$ are modified Bessel functions of the second kind and the parameter $\xi$ is given by

$$\xi = \frac{\omega \rho}{3c} \left( \frac{1}{\gamma^2} + \theta^2 \right)^{3/2} . \tag{7}$$

The term in $K_{2/3}(\xi)$ corresponds to light polarized parallel to the electron orbit plane (horizontally polarized) and the term in $K_{1/3}(\xi)$ corresponds to light polarized perpendicular to the electron orbit plane (vertically polarized). For a discrete monoenergetic electron beam moving through a dipole bending magnet in a storage ring or an accelerator the gyrofrequency is given by

$$\omega_B = \frac{eB}{\gamma m_e c} , \tag{8}$$

where $B$ is the dipole magnetic field strength. The corresponding instantaneous radius of curvature $\rho$ is given by

$$\rho \cong \frac{c}{\omega_B} = \frac{E}{eB}, \tag{9}$$

where $E = \gamma m_e c^2$ is the total electron energy. The critical frequency, defined as the frequency which divides the emitted power in half below and half above is given by

$$\omega_c = \frac{3c\gamma^3}{2\rho} = \frac{3eB}{2m_e c}\left(\frac{E}{m_e c^2}\right)^2 . \tag{10}$$

Thus, the single-particle intensity $I_e(\omega)$ can be written as

$$I_e(\omega) = \frac{3e^2}{4\pi^2 c}\gamma^2 \left(\frac{\omega}{\omega_c}\right)^2 (1+\gamma^2\theta^2)^2 \left[K_{2/3}^2(\xi) + \frac{\gamma^2\theta^2}{1+\gamma^2\theta^2} K_{1/3}^2(\xi)\right] , \tag{11}$$

where $\xi = (\omega/2\omega_c)(1+\gamma^2\theta^2)^{3/2}$.

By neglecting the effects due to the finite size and the angular divergence of the electron beam, and by taking $\theta = 0$ (Kim, 1986),

$$I_e(\omega) = \frac{3e^2}{4\pi^2 c}\gamma^2 H_2(y) , \tag{12}$$

where $H_2(y) = y^2 K_{2/3}^2(y/2)$, with $y = (\omega/\omega_c)$, is called the spectral shape function.

If we consider for example, discrete bunches of electrons as in a laboratory accelerator, to obtain the total energy emitted by an electron bunch per unit of solid angle per unit of frequency, one has to multiply the single-particle intensity by the multiparticle coherent enhancement factor [Schiff 1946, Nodvick & Saxon 1954], thus yielding

$$\frac{d^2W}{d\omega\, d\Omega} = \{N_e[1-f(\omega)] + N_e^2 f(\omega)\} I_e(\omega) , \tag{13}$$

where $N_e$ is the number of electrons in the bunch, $f(\omega)$ is a form factor determined from the spatial electron distribution within the bunch and $I_e(\omega)$ is the single-particle intensity given by Eq. (3) (here with $\vec{r}(\tau)$ standing for the position of the bunch center). The first term in Eq. (13) gives the ISR emission, which scales linearly to the number of electrons. The second term is the CSR emission, which scales as the square of the number of electrons due to the coherent interaction over the frequency range given by the form factor $f(\omega)$.

The size and the shape of the bunch defines the form factor $f(\omega)$, through the Fourier transform of the normalized longitudinal spatial charge distribution of the bunch, $S(z)$:

$$f(\omega) = \left|\int_{-\infty}^{\infty} \exp[i\omega(\hat{n}\cdot\vec{z})/c]\, S(z)\, dz\right|^2 . \tag{14}$$

For a gaussian shape bunch of length $\sigma_b$, the form factor is given by:

$$f(\omega) = \exp[-\omega^2\sigma_b^2/c^2] = \exp[-4\pi^2\sigma_b^2/\lambda^2] , \tag{15}$$

where $\lambda$ is the radiation wavelength at frequency $\omega$. In terms of a time-width defined as $\tau_b = \sigma_b/c$, the form factor can be written as

$$f(\omega) = \exp[-\omega^2 \tau_b^2] \quad . \tag{16}$$

## 4. Simulations in laboratory and flare accelerator scenarios

It should be noted that in a laboratory accelerator, the function S(z) is often assumed to be Gaussian, but in the solar flare accelerator it may be far more complex. The spatial electron distribution can have an effect on the spectral shape of the CSR peak. For instance, simulations using different analytic solutions were tested (Klopf et al. 2007) and showed that good fits are obtained using either Gaussian or hyperbolic secant shape electron bunches for which the form factor $f(\omega)$ is given by,

$$f(\omega) = \text{sech}[\omega \sigma_b / 4c] = \text{sech}[\omega \tau_b / 4] \quad . \tag{17}$$

The hyperbolic secant shape is very similar to the Gaussian shape, but has broader tails. This reduces the sharpness of the CSR peak and improves the fit when compared to the CSR emission from Gaussian shaped bunches. The sparseness of measurements at sub-THz frequencies though makes it difficult to discern anything more than the characteristic bunch size needed to produce a CSR peak at the right frequency, so the simulations here are limited to the *sech* shape.

It is also important to consider that electron beams accelerated in solar flares usually follow a certain energy distribution over a large energy range, unlike the nearly mono-energetic beams typical in laboratory accelerators. Indeed, in the simulations presented here, we consider different bunching structures, electron energy distributions, and assume that only a small fraction of the high-energy electrons have density perturbations sufficient to produce CSR in the solar flare emission. Nonetheless, the $N_e^2$ scaling of the coherent radiation emission produces a striking effect on the spectral emission.

The spectral signature created by the coherent interaction at long wavelengths produces a double peaked spectrum (for example Williams 2002; Carr et al. 2002). An example of this was simulated with the algorithm described above, and is illustrated in Figure 2, which shows the computed ISR/CSR spectrum emitted by discrete compressed electron bunches in the Jefferson Lab Free Electron Laser (FEL) accelerator (Klopf et al. 2007). The spectrum has been computed for a typical accelerator setup in which a monoenergetic beam of electrons ($E$=125 MeV) is compressed into hyperbolic secant shape bunches ($N_e$=8x10$^8$ electrons/bunch) with $\tau_b$=1 ps (or $\sigma_b$=300 µm) and accelerated through a magnetic field with strength $B$=0.1 T. The powerful CSR emission rises from low-frequency up to a cutoff characteristic of the bunch size. At frequencies above the CSR cutoff, the spectral brightness falls until the rising ISR component exceeds the CSR component, above which the ISR component continues to rise up to the characteristic ISR maximum flux set by the electron beam energy and the strength of the magnetic field. Figure 3 shows the ISR/CSR spectra for hyperbolic secant shape bunches with three different time lengths $\tau_b$: 10 ps, 1 ps and 100 fs. A comparative simulation of ISR/CSR spectra for monoenergetic electron beams ($E$=125 MeV) compressed into gaussian and hyperbolic secant shape bunches ($N_e$=8x10$^8$ electrons/bunch) respectively with $\tau_b$=10 ps and $\tau_b$=25 ps and accelerated through a magnetic field with strength $B$=0.1 T is shown in Figure 4. As one can observe, the distinction between the two spectral profiles is seen primarily on the high frequency side of the CSR peak.

Electron beams accelerated in solar flares are usually assumed to follow a power-law energy distribution, instead of being monoenergetic. Sometimes they exhibit an energy break, $E_{break}$, with different spectral indices for energies higher and lower than $E_{break}$ (Lin 2005; Kurt et al. 2010; Holman et al. 2011; Kontar et al. 2011). Several simulations of the ISR/CSR mechanism were done to compare to the November 4, 2003 solar burst data as observed at microwave and sub-THz frequencies (Kaufmann et al. 2004). We have considered normalized single power-law distributions $n(E_e)$ for a total number of electrons $N_e$ with kinetic energies $E_e = E - m_e c^2 = (\gamma - 1) m_e c^2$ within the range from $E_{min}$ to $E_{max}$,

$$n(E_e) = A\, E_e^{-\delta}, \tag{18}$$

where $\delta$ is the spectral index and $A$ is a normalization constant such that

$$\int_{E_{th}}^{E_{max}} n(E_e) dE_e = 1 . \tag{19}$$

Considering that because of Coulomb repulsion microbunching is not possible for electrons with energies below a certain energy threshold $E_{th}$ (Ingelman & Siegbahn 1998), the number of electrons participating in the CSR process, $N_{CSR}$, is just a fraction of the number of high-energy electrons $N_{high}$ ($E_{th} < E_e < E_{max}$),

$$N_{high} = N_e \int_{E_{th}}^{E_{max}} n(E_e) dE_e . \tag{20}$$

The number of low-energy electrons $N_{low}$ ($E_{min} < E_e < E_{th}$) is given by

$$N_{low} = N_e \int_{E_{min}}^{E_{th}} n(E_e) dE_e , \tag{21}$$

One should note that the total number of electrons participating in the ISR process is given by $N_{ISR} = N_{low} + N_{high}(1 - N_{CSR}/N_{high}) = N_e - N_{CSR}$, thus including both low-energy and high-energy electrons. Therefore, the total spectral intensity for the radiating electrons can be written as

$$\frac{d^2 W_{ISR}}{d\omega\, d\Omega} = \frac{d^2 W_{ISR}^{low}}{d\omega\, d\Omega} + \frac{d^2 W_{ISR}^{high}}{d\omega\, d\Omega} + \frac{d^2 W_{CSR}}{d\omega\, d\Omega} . \tag{22}$$

The spectral contributions for the fraction of high-energy electrons participating in the CSR process (those that experience the microbunching instability) and for the high-energy electrons participating only in the ISR process (those that do not microbunch) are given respectively by

$$\frac{d^2 W_{CSR}}{d\omega\, d\Omega} = \left\{ N_{CSR}[1 - f(\omega)] + N_{CSR}^2 f(\omega) \right\} \frac{N_e}{N_{high}} \int_{E_{th}}^{E_{max}} I_e(\omega) n(E_e) dE_e , \tag{23}$$

$$\frac{d^2 W_{ISR}^{high}}{d\omega\, d\Omega} = N_e (1 - N_{CSR}/N_{high}) \int_{E_{th}}^{E_{max}} I_e(\omega) n(E_e) dE_e , \tag{24}$$

where $I_e(\omega)$ is the single-particle intensity for synchrotron emission from highly relativistic electrons given by Eq. (12). The spectral contribution for the low-energy electrons participating in the ISR process is given by

$$\frac{d^2W_{ISR}^{low}}{d\omega\,d\Omega} = N_e \int_{E_{min}}^{E_{th}} I_e^{GS}(\omega) n(E_e) dE_e \;, \quad (25)$$

where $I_e^{GS}(\omega)$ is the single-particle intensity for gyro-synchrotron emission from mildly relativistic electrons (Ramaty 1969), since we consider an energy threshold $E_{th}$ of a few MeV.

The spectral flux densities (power per unit frequency per unit area) for the high-energy electrons are obtained from the spectral intensities given by Eqs. (23) and (24) through a normalization factor $\eta$ which accounts for the effects due to finite emittance. The spectral flux density for the low-energy electrons is obtained through the solution of the transfer equation for an homogeneous source using Ramaty's gyrosynchrotron/synchrotron algorithm (Ramaty 1969; Ramaty et al. 1994), with the emissivity and absorption coefficient evaluated from the spectral intensity given by Eq. (25). The total spectral flux density must then be converted into solar flux units (1 SFU = $10^{-22}$ W m$^{-2}$ Hz$^{-1}$). In our calculations we assume a total number of electrons, $N_e = N_{low} + N_{high} = N_{ISR} + N_{CSR}$, plausible for solar flares and set the normalization factor $\eta$ to scale the ISR emission. The fraction of high-energy electrons participating in the CSR process, $N_{CSR}/N_{high}$, is set to scale the peak of the CSR emission and the bunch length $\tau_b$ is set to determine the peak frequency of the CSR emission.

In Figure 5 the November 4, 2003 solar flare microwave data for the two burst event time structures shown in Figure 1a, P$_1$ at 19:44:00 UT and P$_4$ at 19:48:20 UT, are fitted using hyperbolic secant shape bunches and a single power-law electron distribution with $E_{min}$=50 keV, $E_{max}$=100 MeV, $E_{th}$=5 MeV and spectral index $\delta$=2. In the calculation of the spectral flux density for the low-energy electrons we have assumed a source size $\Omega = 20''$. For time structure P$_1$, $N_e=10^{32}$, $N_{low}$=9.9022x10$^{31}$, $N_{high}$=9.78x10$^{29}$, $N_{CSR}$=4.580x10$^{-15}$ $N_{high}$ and $\tau_b$=30 ps (9 mm). For time structure P$_4$, $N_e$=5x10$^{31}$, $N_{low}$=4.9511x10$^{31}$, $N_{high}$=4.89 x10$^{29}$, $N_{CSR}$=1.117x10$^{-14}$ $N_{high}$ and $\tau_b$=47 ps (14 mm). In Figures 6 and 7 we show the contributions to the total flux respectively from the ISR and the CSR processes and from the low-energy and the high-energy electrons, corresponding to the simulations displayed in Figure 5 for time structures P$_1$ and P$_4$.

In Figure 8 we compare the fits to the November 4, 2003 solar flare microwave using single power-law distributions with distinct values for $E_{th}$ and the same parameters $E_{min}$=50 keV, $E_{max}$=100 MeV and spectral index $\delta$=2. For all values of $E_{th}$, we have used hyperbolic secant shape bunches with $\tau_b$=30 ps for time structure P$_1$ and $\tau_b$=47 ps for time structure P$_4$. As one can observe, the spectral index of the radiation in the sub-THz range of frequencies increases with $E_{th}$ (approaching an asymptotic value in the limit $E_{th} \rightarrow E_{max}$), such that a better fit to the November 4, 2003 solar flare data from the SST can be obtained. In Table 1 we list the number of low-energy electrons, $N_{low}$, the number of high-energy electrons, $N_{high}$, and the number of high-energy electrons participating in the CSR process, $N_{CSR}$, required to fit the observed fluxes for time structures P$_1$ and P$_4$, corresponding to the plots shown in Figure 8.

In Figure 9 we show the fits to the November 4, 2003 solar flare microwave data for time structures P$_1$ and P$_4$ using single power-law distributions with distinct values for the spectral index and the same parameters $E_{min}$=50 keV, $E_{max}$=100 MeV and $E_{th}$=5 MeV. As one can observe, a satisfactory fit to the sub-THz data from the SST can be obtained for all values of the spectral index δ considered in the simulations. However, a spectral index δ < 3 is required in order to obtain an ISR component with flux that keeps increasing for larger frequencies and maximizes in the THz range. In Table 2 we list the number of low-energy electrons, $N_{low}$, the number of high-energy electrons, $N_{high}$, and the number of high-energy electrons participating in the CSR process, $N_{CSR}$, required to fit the observed fluxes for time structures $P_1$ and $P_4$, corresponding to the plots shown in Figure 9.

## 5. Discussion

We have shown that simulations based on the ISR/CSR mechanism in laboratory accelerators can reproduce well the observed solar flare double-spectra. The high-frequency component peaking in the THz range is produced by ISR from high-energy electrons and the microwave component peaking in the GHz range is produced simultaneously by ISR from low-energy electrons and CSR from a small fraction of the high-energy electrons, the latter as the result of microbunching instability.

The formation of bunches in a beam of high energy accelerated electrons is more likely to happen at locations in the flaring region where the magnetic field is more intense and exhibiting more complex topology. It shall also be more efficient in denser beams. This suggest that the mechanism might be more effective deeper in the solar atmosphere, at magnetic structures closer to the sunspots, rather than higher in the corona, such as at the loop tops.

The ISR/CSR mechanism is highly efficient and might be present in every flare. The relative importance of flux produced by CSR compared to other mechanisms, in particular to flux due to gyro-synchrotron from mildly relativistic electrons, depends on the form factor for the CSR/ISR process and the relative number of electrons undergoing microbunching and emitting CSR. These parameters depend on the magnetic field strength, the spatial complexity, and the beam density, which may be more ideal for microbunching and CSR emission closer to the Sun's surface.

However, observation of CSR emission may not always be possible. The plasma densities surrounding the accelerated beam of electrons must have a plasma frequency that corresponds to or is smaller than the critical frequencies allowing the CSR microwaves to escape and be observable. This requirement is in fact the same for microwaves originated from other mechanisms such as the gyrosynchrotron emission. For typical microwave emissions peaking around 10 GHz, the plasma densities in active regions where the flare radiation originates must be smaller than about $10^{12}$ cm$^{-3}$ at the lower chromosphere. Furthermore in the deep atmosphere, the plasma parameter (ratio of plasma to magnetic pressures) is >> 1, whereas it is << 1 in the corona. Therefore, we do expect stronger magnetic field inhomogeneities in the lower atmosphere, compared to the middle/high corona where the magnetic pressure overcomes the plasma pressure. Emission from lower locations is also consistent with a classical flare picture model where the flare is triggered by small and compact emerging loops at the low atmosphere.

The high frequency radiation alone might be explained by a number of distinct emission mechanisms, including the ISR, which may be acting at the same time, with different proportions (Fleishman & Kontar 2010). The CSR/ISR emission from a microbunching instability might be present at every flare, with different relative intensity of CSR in comparison to other mechanisms producing microwaves. Their distinction is a subject for further investigation, both by theory and by observations.

The simulations presented here were compared to the November 4, 2003 burst data, observed at two frequencies by the SST submillimeter telescope, and at microwaves by Owens Valley Solar Array and Itapetinga 13.-7 telescope (44 GHz for time structure P4). Successful simulations were shown to become possible for a number of assumed electron energy distributions in the beam. We have analyzed monoenergetic and power-law distributions. Both can reproduce the observed solar flare data. The monoenergetic distribution is usually the condition found in laboratory accelerators, but is unlikely to occur in solar flare accelerator.

More attention has been given to power-law electron energy distributions, which are usually assumed for solar flare accelerators. Successful simulations are obtained for single power-law electron distributions considering that the CSR process becomes possible only for electron energies above a threshold $E_{th}$, set at sufficiently high energies (> 2 MeV), below which microbunches cannot be formed due to Coulomb repulsion (Ingelman & Siegbahn 1998). The comparison between simulated and observed fluxes is excellent. The steeper index for data at frequencies smaller than 10 GHz may be explained by the Razin suppression of synchrotron radiation by ambient plasma (Ginzburg and Syrovatskii 1965; Ramaty & Lingenfelter 1967), a condition that is not encountered in laboratory accelerators, and not taken into account in the simulation algorithm implemented here.

The Razin suppression depends on the density and magnetic field of the media surrounding the acceleration site and the accelerated beam propagation path. The effect is negligible on ISR emission in the sub-THz or larger frequencies, because it would require plasma densities unrealistically high at the solar atmosphere for typical magnetic fields B in the range of $10^2$ to $10^3$ Gauss. However it becomes important for the CSR at lower microwave frequencies, smaller than 5 GHz. For the same range of B, it should require densities found in the solar corona. In Figure 5 and following ones the observed emission at lower microwaves frequencies exhibit a spectral index steeper compared to the predicted by simulations (which do not include the Razin effect).

Simulations show that the contribution to the microwave spectral component produced by ISR from the low-energy electrons (those with energy below the threshold), which represents most of the accelerated population, is about one order of magnitude weaker than the contribution produced by CSR from the high-energy electrons due to the microbunching instability. The value of the electron low energy cutoff level is not critical in the ISR/CSR calculations here. The reduction in $E_{min}$ only slightly changes the microwave spectral component. Furthermore, it has been found that the ISR/CSR mechanism is extremely efficient: only a very small fraction of the high-energy electrons in the beam (about $10^{-15}$ $N_{high}$) is required to explain the microwave radiation as coherent synchrotron generated by this mechanism (for a total number of electrons $N_e = N_{low} + N_{high} > 10^{31}$).

The spectral emission profiles computed through the simulations of the ISR/CSR mechanism presented here depend, of course, on the interplay between the parameters which define the microbunch structure and the electron energy distribution, as well as on the magnetic field strength. In particular the high-frequency spectral component peaking in the THz range, which is produced by ISR from high-energy electrons, depends only on the parameters of the electron energy distribution and the magnetic field strength. Using a single power-law distribution, for an energy range from 50 keV to 100 MeV, a spectral index index $\delta < 3$ is required in to obtain an increasing ISR spectrum peaking in the THz range of frequencies that fits the November 4, 2003 solar flare data from the SST (0.2 and 0.4 THz). Such an upper limit for the spectral index is not consistent with the values in the range from 3 to 7 which are usually considered in standard calculations of gyrosynchrotron/synchrotron emission from non-thermal power-law relativistic electrons (Dulk & Marsh 1982; Dulk 1985). However, it is important to note that such calculations are based on formulas intended to describe the well-known microwave spectral component peaking in the GHz range of frequencies, which are derived for power-law electron distributions in the "mild" energy range from 10 keV to 1 MeV and takes into account the gyro-synchrotron and synchrotron processes as well as the effects of the medium and self-absorption (Ramaty & Lingenfelter 1967, Ramaty 1969, Dulk 1985). On the other hand, in our simulations of the ISR/CSR mechanism, a much higher upper limit for the electron energies must be considered in order to obtain an increasing high-frequency component peaking in the THz range, that is produced by ISR from high-energy electrons (one should recall that the microwave spectral component peaking in the GHz range is produced mostly by a fraction of the high-energy electrons through the CSR process). Nevertheless, by reducing the upper energy limit $E_{max}$ to a few MeV the simulations developed here are suggestive but inadequate for frequencies lower than the microwave peak.

Although reasonable for the high-energy electrons in the power-law distribution, the spectral index $\delta < 3$ required to fit the sub-THz data from the SST cannot be regarded as realistic for the entire energy range from 50 keV to 100 MeV. The results of simulations which will be presented in a forthcoming article suggest that a double power-law distribution with an index break point $E_{break}$ of a few MeV and spectral indices $\delta \geq 3$ and $\delta \leq 2$ respectively below and above $E_{break}$ is most likely to provide a consistent description of the "double spectral" feature observed for several solar flares based on the ISR/CSR microbunching mechanism. Assuming that most of the hard X- and gamma-ray radiation is produced by accelerated electron collisions at denser regions of the solar atmosphere, the observed photon energy spectrum is proportional to the accelerated electron energy distribution (Tandberg-Hanssen & Emslie 1988). Indeed, the observed hard X- and gamma-ray time profiles for the November 4, 2003 burst are almost identical to the two sub-THz profiles shown in Figure 1a, with a photon energy index break above 10 MeV (Kurt et al. 2010).

## 5. Concluding remarks

High-energy electrons accelerated in solar flares produce incoherent synchrotron radiation (ISR). The beams may undergo density perturbations under conditions for microbunching instability thus producing intense broadband coherent synchrotron radiation (CSR) at lower frequencies with wavelength scales of the order or longer than the microbunch size. This mechanism can explain the "double-spectral" structures observed for several solar bursts: one component due to ISR, maximizing in the THz range of frequencies, and one due to the simultaneous broadband CSR, maximizing in the GHz range. The instability arises by interaction of ISR photons with the accelerated electrons and/or by magnetic field irregularities in the medium where the beam propagates (Venturini & Warnock 2002; Stupakov & Heifets 2002). There are several solar flare scenarios giving such conditions. In the complex fine magnetic structures of solar active regions it is not difficult to conceive the presence of spatial inhomogeneities in the magnetic flux traversed by the accelerated beam of electrons, such as highly sheared fields in multiple loops, closing on double or quadruple poles on the photosphere, with fine helical structures in space (Sturrock 1987; Antiochos 1998). These complex topologies are supported by observations in the visible, infrared, UV and hard X-rays (Hanaoka 1999; Gary & Moore 2004; Socas-Navarro 2005).

Simulations of the ISR/CSR mechanism have been carried out for different electron energy distributions: monoenergetic and power-law. They have been successfully applied to the large solar burst of November 4, 2003, observed from radio to gamma-ray wavelengths, fitting particularly well for power-law distributions considering that electrons with energies smaller than a threshold of 2 MeV or greater do not participate in the CSR process because the Coulomb repulsion prevents the bunching. Nevertheless, the milder energy electrons produce their own ISR in the microwave range of frequencies, with intensities about one order of magnitude smaller that the CSR intensities.. The CSR emission from even a very localized region of a flare exhibiting microbunching is extremely efficient. Only a very small fraction of electrons in the beam is required to produce the enhancement of the microwave spectral component due to the CSR process (about $10^{-15}$ times the total number of electrons in the beam with energies above the threshold). In our simulations of the ISR/CSR mechanism using a single-power law distribution, we have assumed a total number of electrons a few orders of magnitude smaller than the number usually assumed for larger flares ($10^{34}$).

This interpretation has no conflict with the well known solar burst microwave spectral and temporal features. It is plausible to conceive the contribution of a very small portion of high-energy electrons, producing ISR, that also contribute for CSR at microwaves at least for certain flares, to explain the "double spectra" in frequency. However for models assuming the acceleration sites deep into the solar atmosphere, the high plasma densities may prevent microwaves to escape, irrespectively from the mechanism they are produced (as for example Sakai et al. 2006 model explaining the sub-THz emission as Langmuir waves).

The ISR/CSR mechanism presented here gives one alternative explanation to the well known question on the electron number discrepancy by orders of magnitude when comparing numbers as derived from X-rays or from microwaves, respectively (Brown 1971; Brown and Melrose 1977; Dulk and Marsh 1982; Kai 1986). One known explanation has been suggested with assumptions of homogeneous sources, thick target collision condition and weaker magnetic fields (hundreds of Gauss, as found in the solar corona) (Gary 1985). However certain large bursts, such as the example selected here, the electron number derived from gyro-synchrotron at microwaves exceeds largely the number derived from hard X-rays (White et al. 2003, Raulin et al. 2004). These microwaves are easier explainable by CSR produced by microbuncing in the ISR radiating electron beam. When the ISR/CSR mechanism is considered to explain the microwave spectral component, the total number of accelerated electrons involved should be derived from the ISR spectrum rather than from the microwave spectrum that might be dominated by CSR and/or by ISR from electrons with milder energies, smaller than the minimum required to obtain bunching, described by single or double power-law distributions. Using the Ramaty gyrosynchrotron/synchrotron algorithm (Ramaty 1969; Ramaty et al. 1994) we may assume the ISR spectral maximum in the 1-10 THz range and extrapolate prediction of significant fluxes in the hard X- and γ-ray ranges, comparable to Coronas observations (Myagkova et al. 2004; Kurt et al. 2010) within an order of magnitude.

It is important to emphasize that a more complete spectral coverage in the THz range of frequencies is necessary in order to provide better observational constraints on the ranges of physical parameters involved in the ISR/CSR mechanism presented here as well as in the other several emission mechanisms suggested to explain the new THz spectral structure feature. Since the terrestrial atmosphere is opaque to almost the whole THz frequency range, this requires new observations carried out from space. Solar activity can also be observed from the ground through few atmospheric THz transmission "windows" at exceptionally good high altitude locations (Lawrence 2004; Suen, Fang & Lubin 2014). Experiments SIRE (Deming, Kostiuk & Glenar 1991) and DESIR (Trottet et al. 2006) have been proposed to observe solar flares in the THz range from satellites. The SOLAR-T solar flare experiment, carrying telescopes at 3 and 7 THz, has been recently completed, to be flown on long duration stratospheric balloons missions (Kaufmann et al. 2014).

**Acknowledgements**: The authors acknowledge the referee comments which have improved the article presentation. These researches were partially supported by Brazilian agencies FAPESP (Contract no. 2006/06847-1), CNPq, Mackpesquisa, US AFOSR, and by US Jefferson Lab (DOE Contract # DE-AC05-06OR23177).

**Captions to the figures**

**Figure 1.** (a) The time profile of the November 4, 3003 solar burst as observed by SST at 0.2 and 0.4 THz, and by Owens Valley Solar Observatory at 15.6 GHz. (b) Spectra for two major peaks $P_1$ and $P_4$ showing the double-structure with a minimum at about 100 GHz.

**Figure 2.** Simulation for electron beam parameters typical for the Jefferson Lab FEL accelerator: ISR/CSR spectrum for monoenergetic electron beams ($E_e = 125$ MeV) compressed into hyperbolic secant shape bunches ($N_e = 8 \times 10^8$ electrons/bunch) with $\tau_b = 1$ ps, accelerated in a dipole magnet (B = 0.1 T).

**Figure 3.** ISR/CSR spectra for monoenergetic electron beams ($E_e = 125$ MeV) compressed into hyperbolic secant shape bunches ($N_e = 8 \times 10^8$ electrons/bunch) with $\tau_b$ = 10 ps (dashed), 1 ps (dot-dashed), 100 fs (dotted) accelerated through a magnetic field with strength $B = 0.1$ T; ISR spectrum only (solid).

**Figure 4.** Comparative simulation of ISR/CSR spectra for monoenergetic electron beams ($E_e$=125 MeV) compressed into gaussian shape (solid) and hyperbolic secant shape (dashed) bunches ($N_e = 8\times10^8$ electrons/bunch) with $\tau_b = 10$ ps and $\tau_b = 25$ ps respectively, accelerated through a magnetic field with strength $B = 0.1$ T.

**Figure 5.** Fit to the November 4, 2003 solar flare microwave data from the Owens Valley Solar Array (OVSA) and the Solar Submillimeter Telescope (SST) for the two burst event time structures shown in Figure 1a: $P_1$ at 19:44:00 UT (circles) and $P_4$ at 19:48:20 UT (squares). We have used hyperbolic secant shape bunches and a single power-law electron distribution with $E_{min}$=50 keV, $E_{max}$=100 MeV, $E_{th}$=5 MeV and spectral index $\delta$=2. In the calculation of the spectral flux density for the low-energy electrons ($E_{min} < E < E_{th}$) we have considered a source with size $\Omega = 20''$. For time structure $P_1$, $N_e$=$10^{32}$, $N_{low}$=9.902x$10^{31}$, $N_{high}$=9.78x$10^{29}$, $N_{CSR}$=4.58x$10^{-15}$ $N_{high}$ and $\tau_b$=30 ps (9 mm). For time structure $P_4$, $N_e$=5x$10^{31}$, $N_{low}$=4.951x$10^{31}$, $N_{high}$=4.89x$10^{29}$, $N_{CSR}$=1.117x$10^{-14}$ $N_{high}$ and $\tau_b$=47 ps (14 mm).

**Figure 6.** Contributions to the total flux from the ISR and the CSR processes, corresponding to the simulations displayed in Figure 5 for time structures $P_1$ (a) and $P_4$ (b).

**Figure 7.** Contributions to the total flux from the low-energy electrons ($E_{min} < E < E_{th}$) and the high-energy electrons ($E_{th} < E < E_{max}$), corresponding to the simulations displayed in Figure 5 for time structures $P_1$ (a) and $P_4$ (b).

**Figure 8.** Comparison between the fits to the November 4, 2003 solar flare microwave data for time structures $P_1$ (a) and $P_4$ (b) obtained using single power-law electron distributions with distinct values for the threshold energy $E_{th}$ and the same parameters $E_{min}$ =50 keV, $E_{max}$ = 100 MeV and spectral index $\delta$ = 2.

**Figure 9.** Comparison between the fits to the November 4, 2003 solar flare microwave data for time structures $P_1$ (a) and $P_4$ (b) obtained using single power-law distributions with distinct values for the spectral index $\delta$ and the same parameters $E_{min}$ = 50 keV, $E_{max}$ = 100 MeV and $E_{th}$ = 5 MeV.

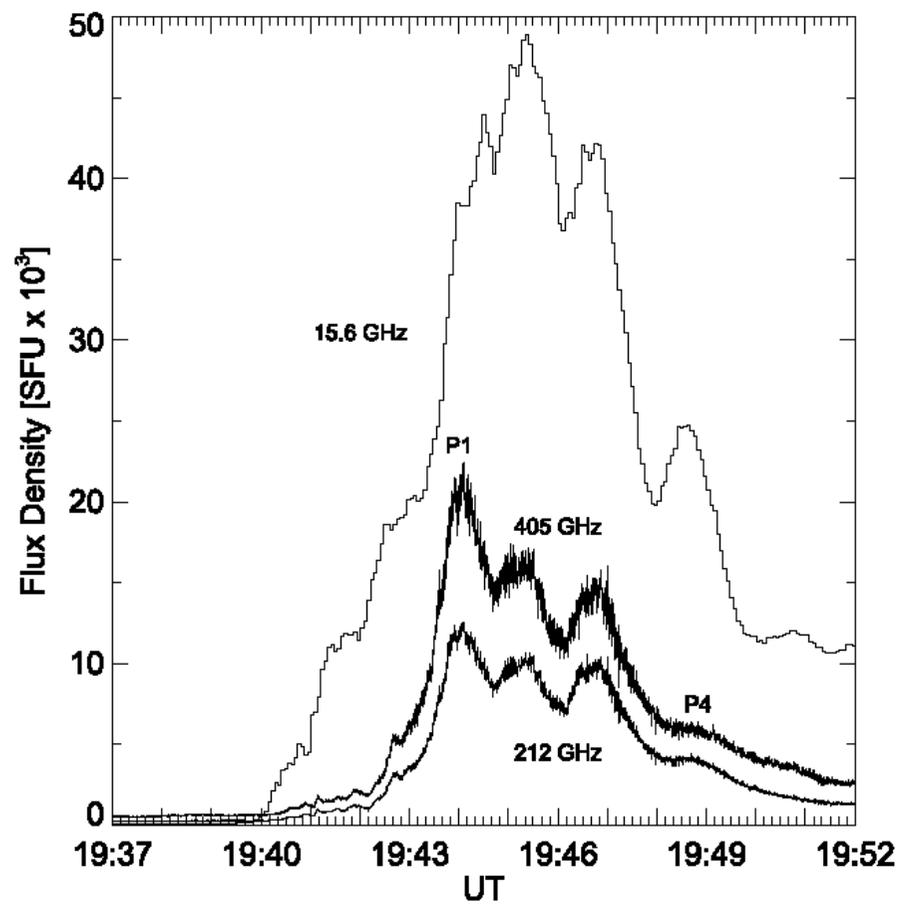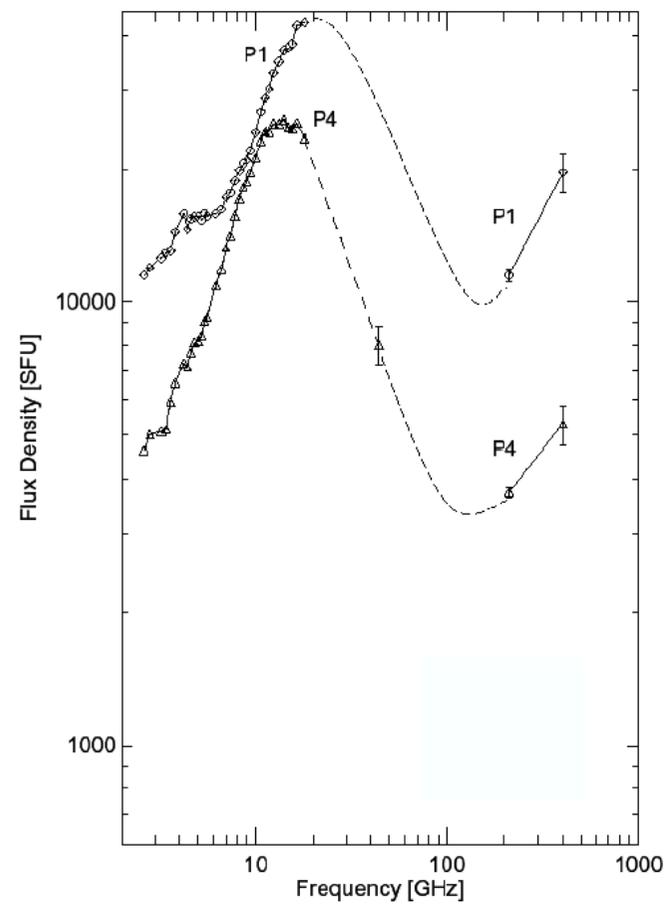

(a)  (b)

Figure 1 (ab)

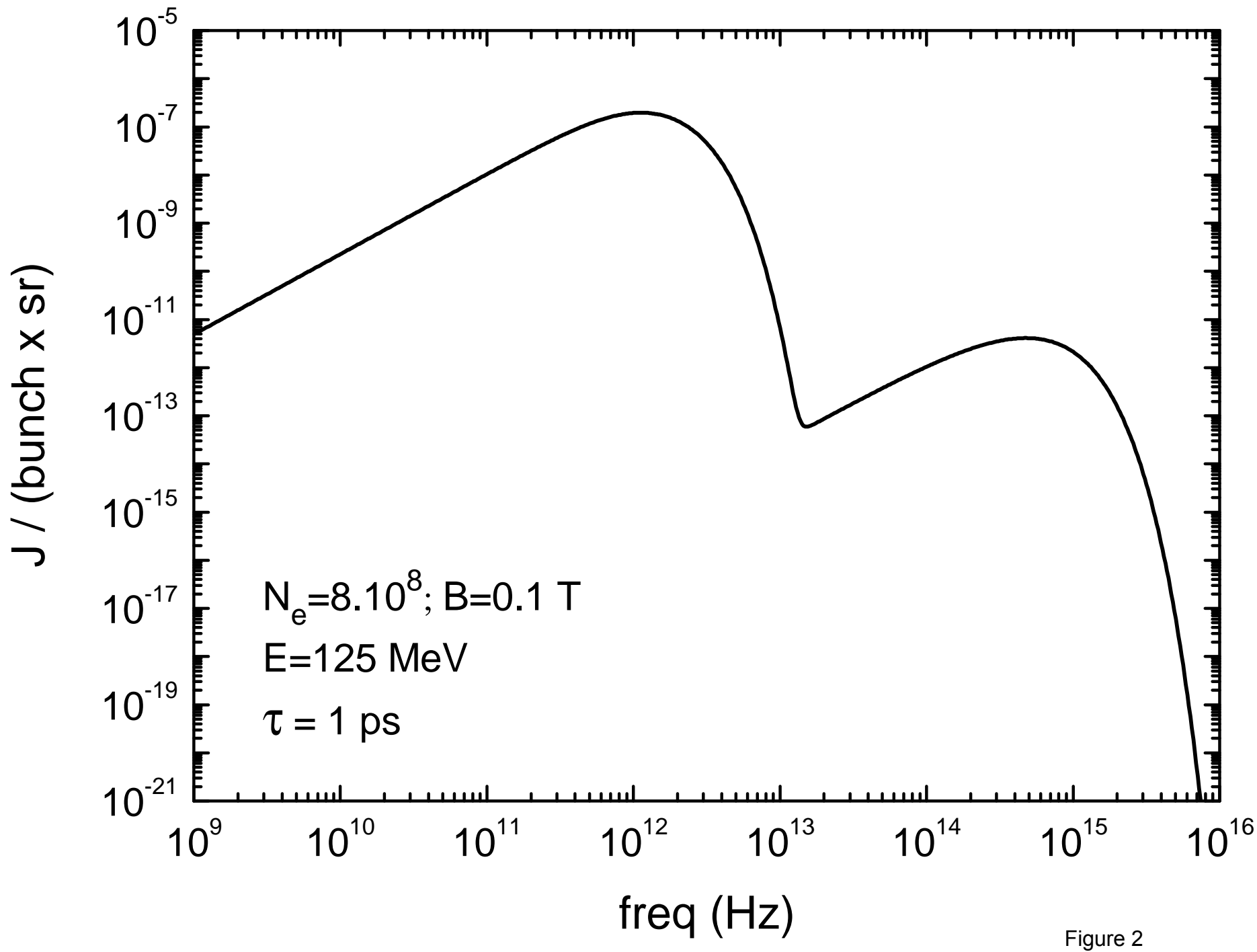

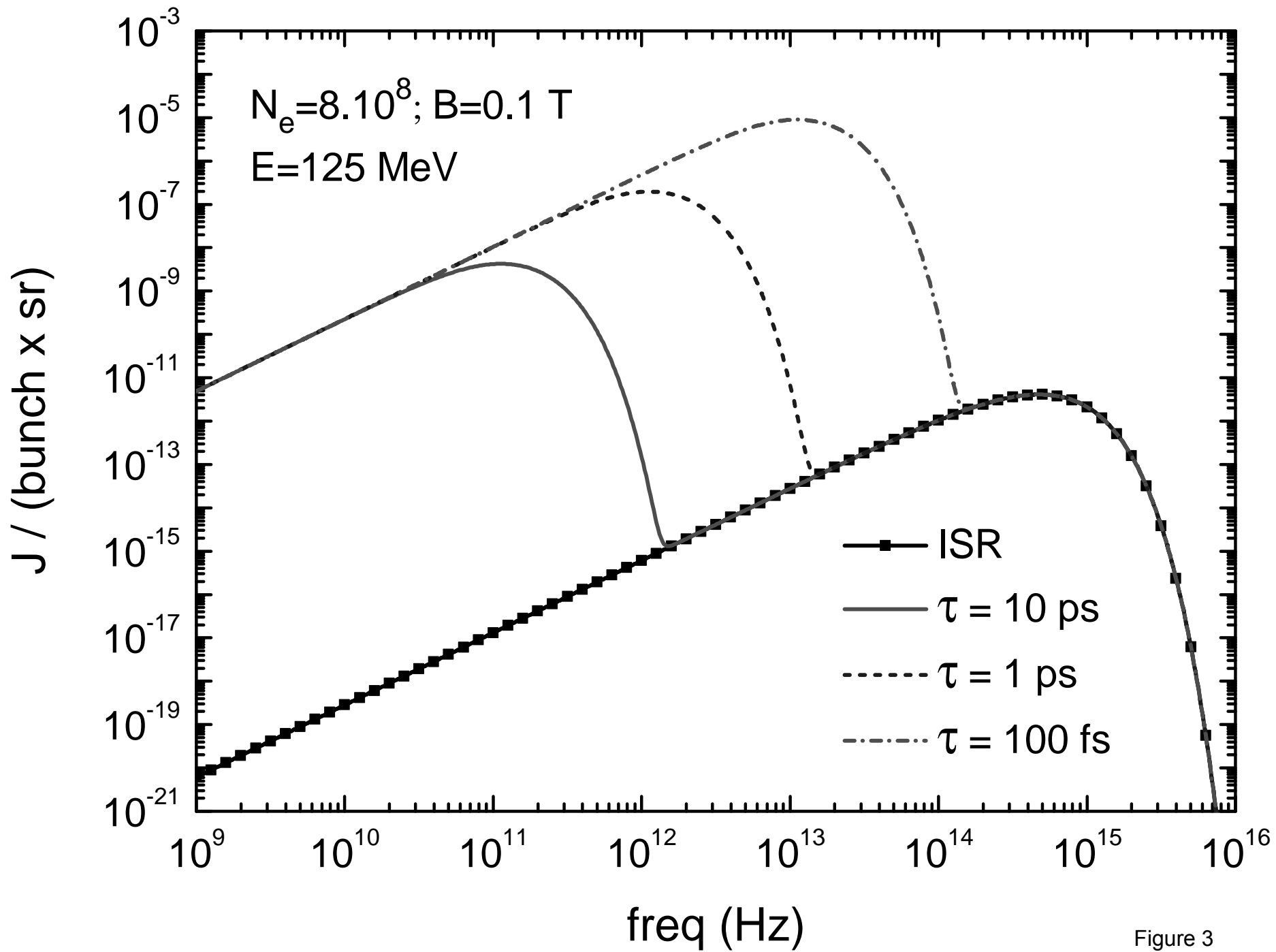

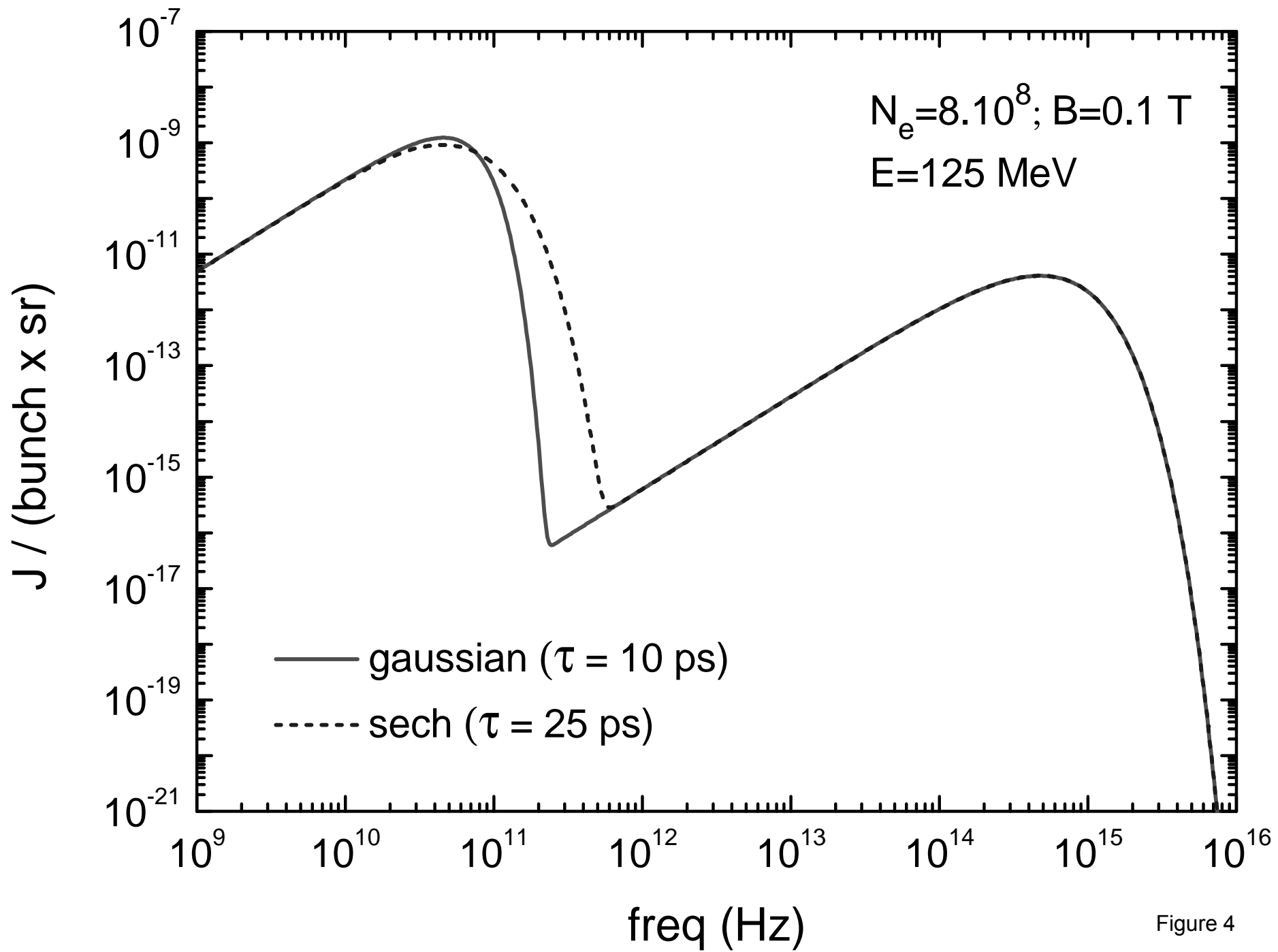

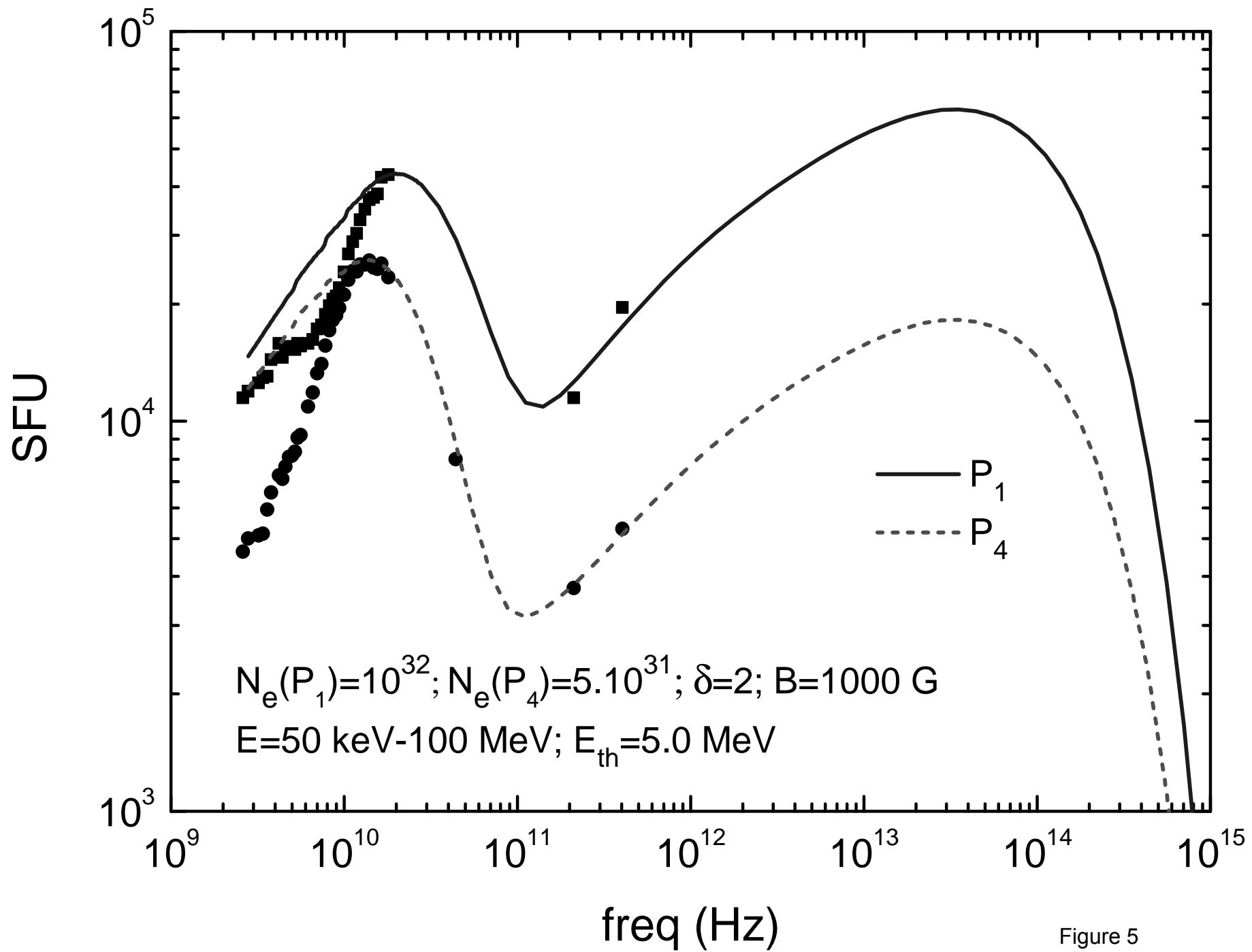

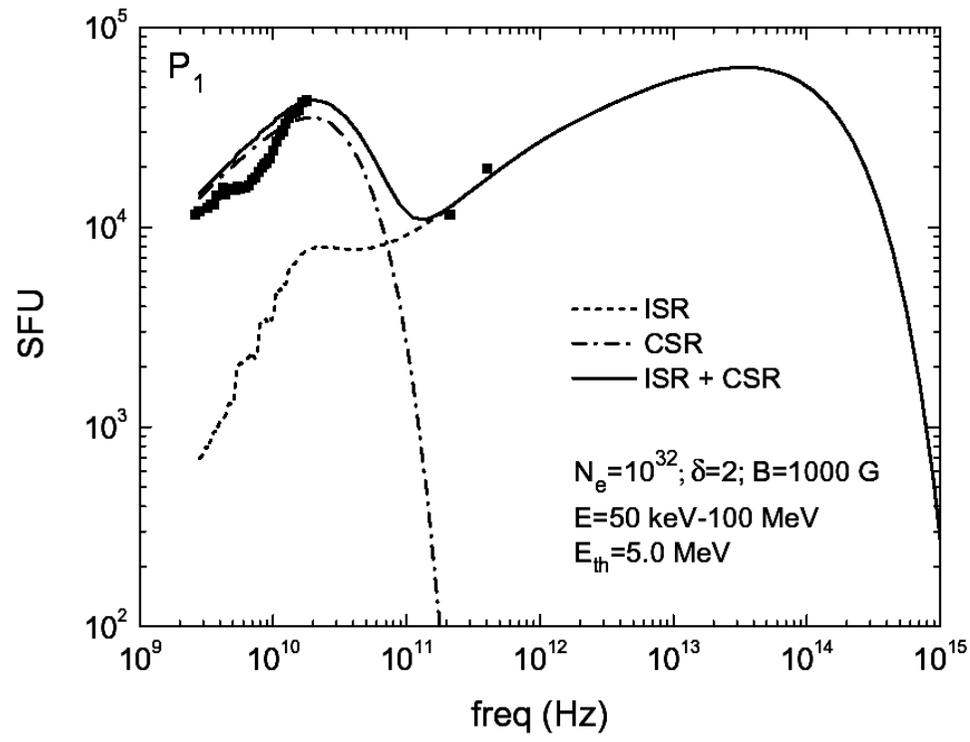 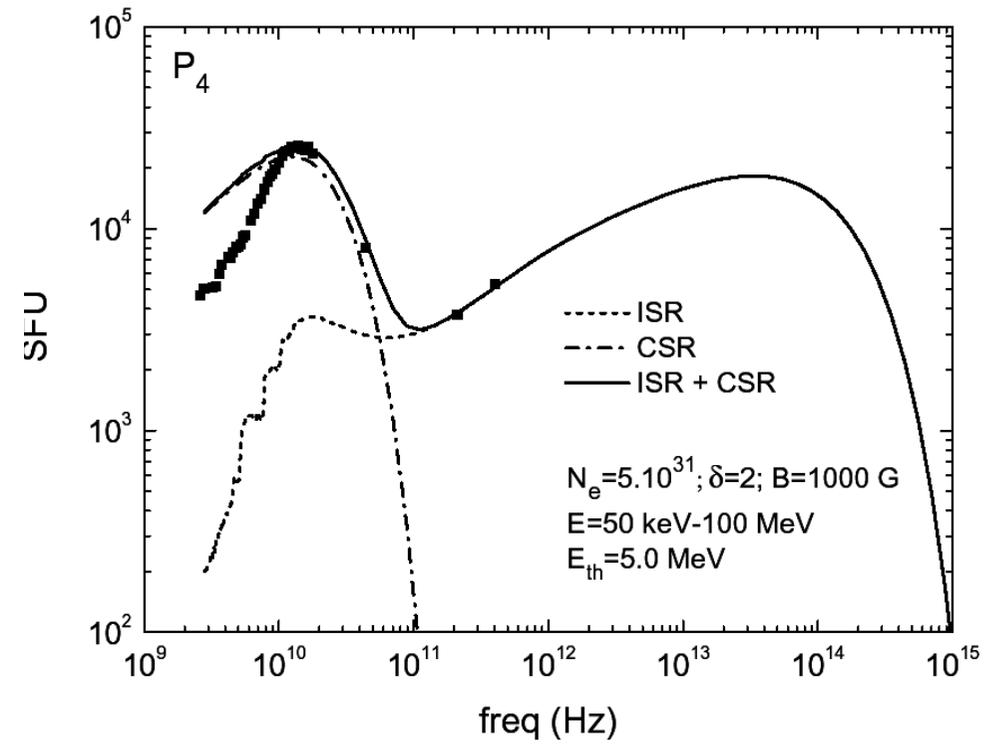

Figure 6 (ab)

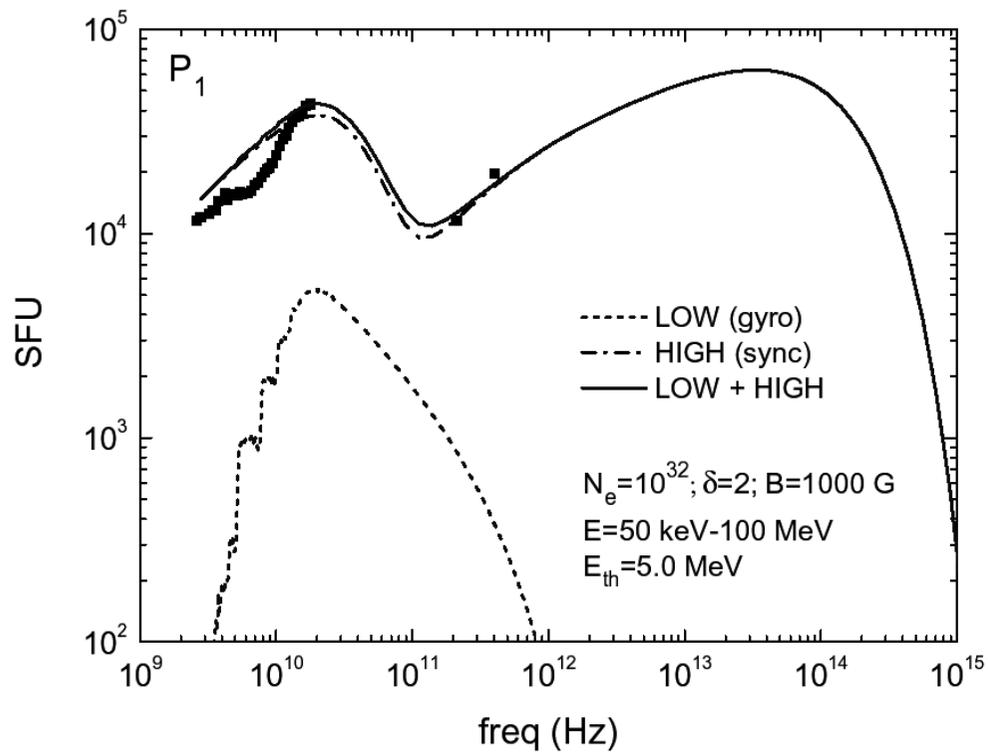 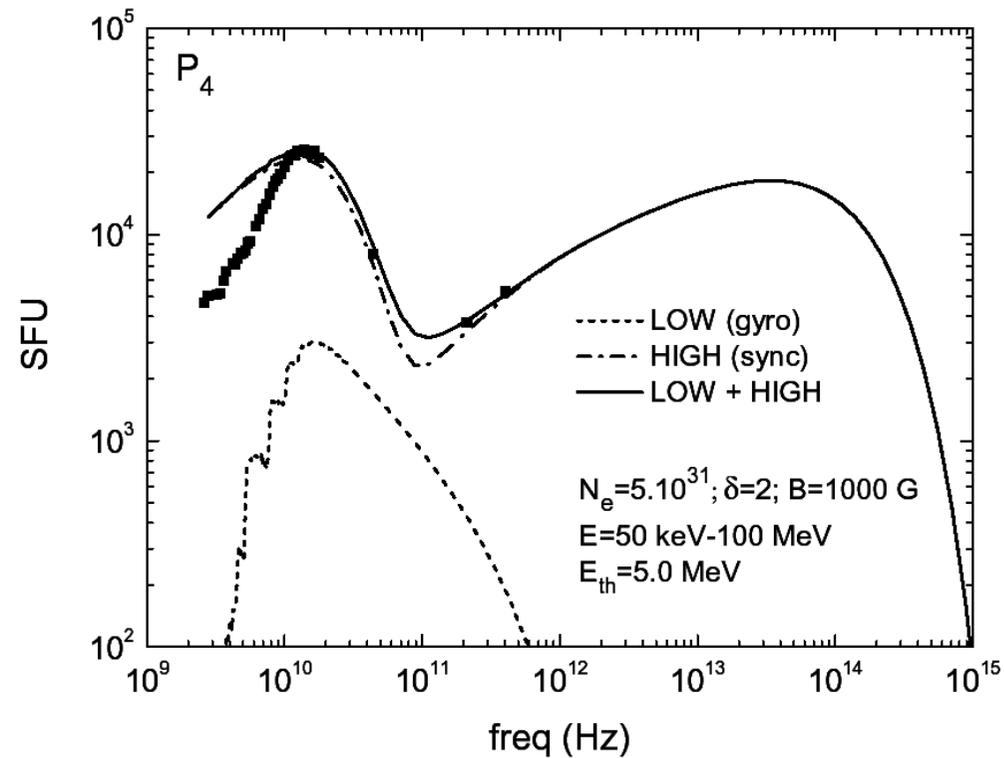

(a)            (b)

Figure 7 ab)

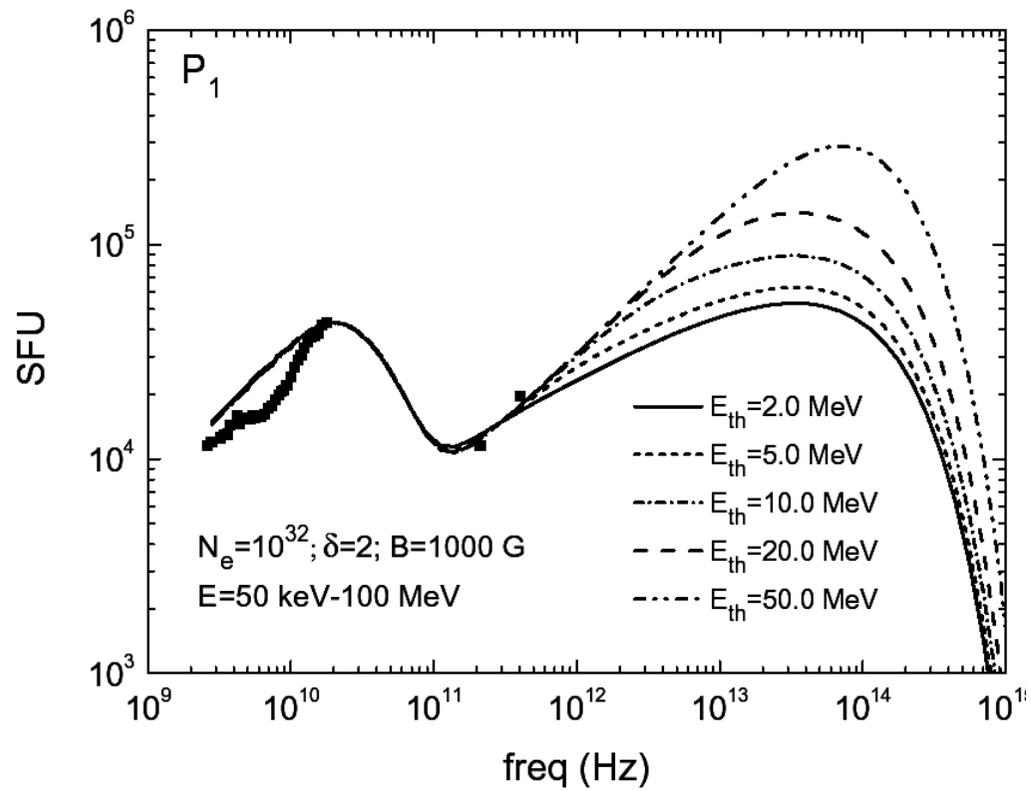 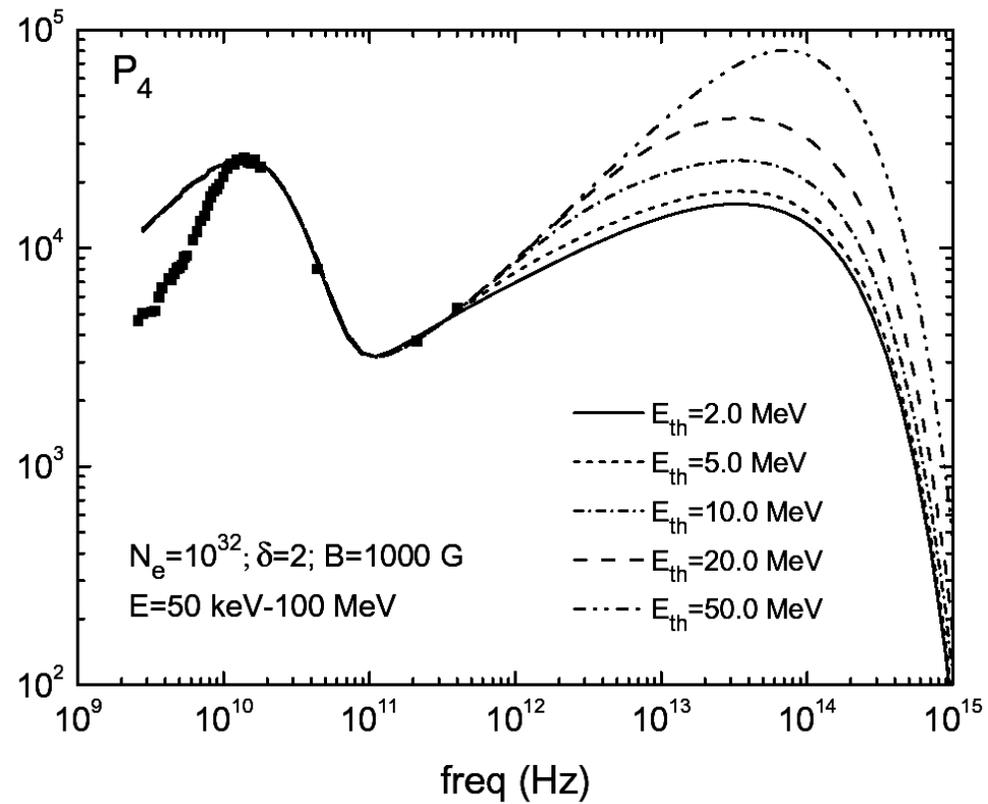

Figure 8 (ab)

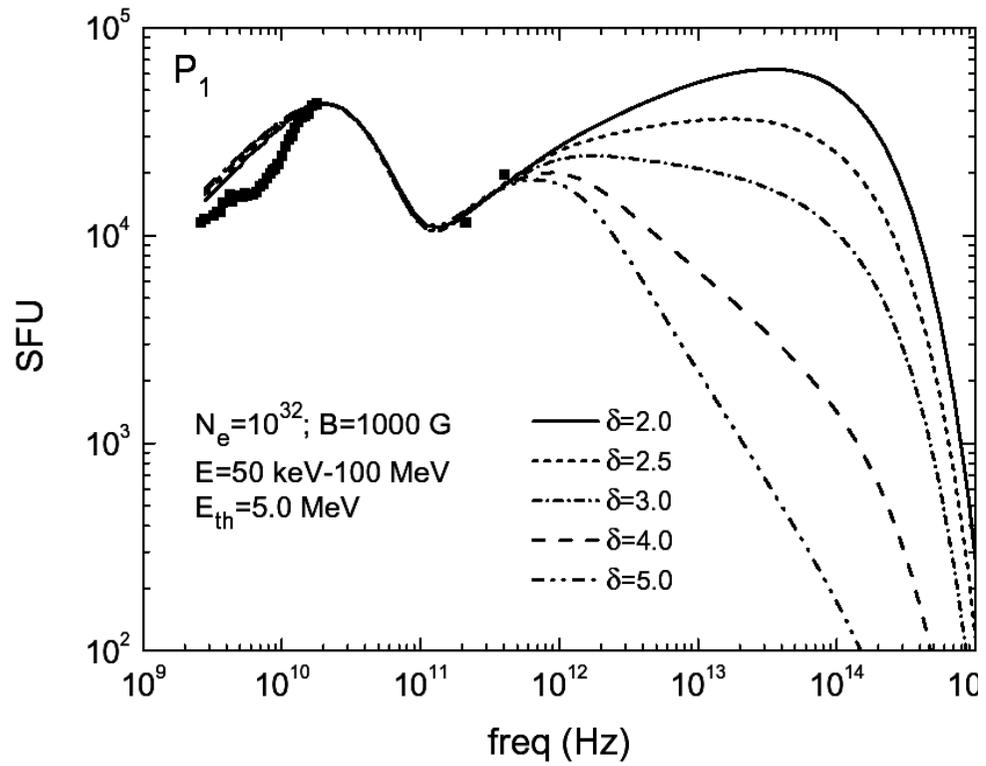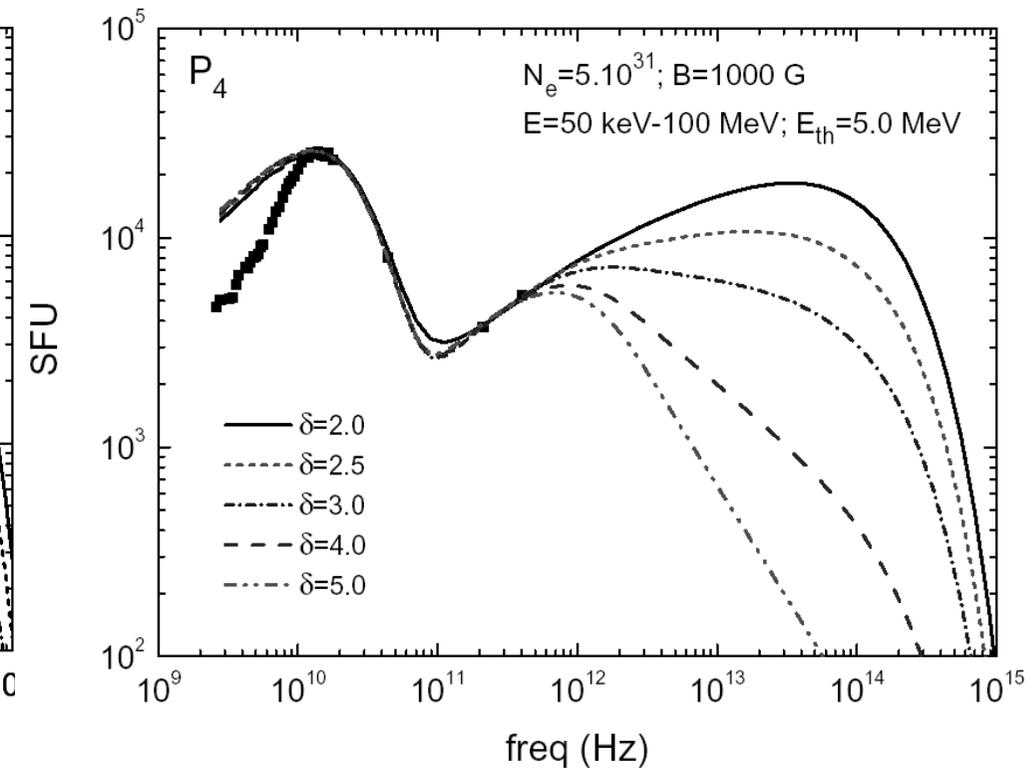

(a)                          (b)

Figure 9(ab)

| $E_{th}$ (MeV) | P$_1$ | | | P$_4$ | | |
| --- | --- | --- | --- | --- | --- | --- |
| | $N_{low}$ | $N_{high}$ | $N_{CSR}$ | $N_{low}$ | $N_{high}$ | $N_{CSR}$ |
| 2.0 | 9.765x10$^{31}$ | 2.354x10$^{30}$ | 6.071x10$^{15}$ | 4.882x10$^{31}$ | 1.177x10$^{30}$ | 7.269x10$^{15}$ |
| 5.0 | 9.902x10$^{31}$ | 9.775x10$^{29}$ | 4.460x10$^{15}$ | 4.951x10$^{31}$ | 4.888x10$^{29}$ | 5.436x10$^{15}$ |
| 10.0 | 9.956x10$^{31}$ | 4.437x10$^{29}$ | 3.139x10$^{15}$ | 4.978x10$^{31}$ | 2.219x10$^{29}$ | 3.851x10$^{15}$ |
| 20.0 | 9.980x10$^{31}$ | 1.990x10$^{29}$ | 2.121x10$^{15}$ | 4.990x10$^{31}$ | 9.952x10$^{28}$ | 2.621x10$^{15}$ |
| 50.0 | 9.995x10$^{31}$ | 4.99x10$^{28}$ | 1.068x10$^{15}$ | 4.997x10$^{31}$ | 2.499x10$^{28}$ | 1.324x10$^{15}$ |

Table 1

| δ | P$_1$ | | | P$_4$ | | |
| --- | --- | --- | --- | --- | --- | --- |
| | $N_{low}$ | $N_{high}$ | $N_{CSR}$ | $N_{low}$ | $N_{high}$ | $N_{CSR}$ |
| 2.0 | 9.902x10$^{31}$ | 9.775x10$^{29}$ | 4.460x10$^{15}$ | 4.951x10$^{31}$ | 4.888x10$^{29}$ | 5.436x10$^{15}$ |
| 2.5 | 9.989x10$^{31}$ | 1.029x10$^{29}$ | 1.439x10$^{15}$ | 4.995x10$^{31}$ | 5.148x10$^{28}$ | 1.727x10$^{15}$ |
| 3.0 | 9.998x10$^{31}$ | 1.052x10$^{28}$ | 4.593x10$^{14}$ | 4.999x10$^{31}$ | 5.260x10$^{27}$ | 5.460x10$^{14}$ |
| 4.0 | 9.999x10$^{31}$ | 1.082x10$^{26}$ | 4.562x10$^{13}$ | 4.999x10$^{31}$ | 5.411x10$^{25}$ | 5.445x10$^{13}$ |
| 5.0 | 9.999x10$^{31}$ | 1.110x10$^{24}$ | 4.543x10$^{12}$ | 4.999x10$^{31}$ | 5.552x10$^{23}$ | 5.442x10$^{12}$ |

Table 2